\documentclass[conference,compsoc]{IEEEtran}

\usepackage[utf8]{inputenc}
\usepackage[english]{babel}
\usepackage[T1]{fontenc}
\usepackage{amsmath}
\usepackage{amsthm}
\usepackage{xspace}
\usepackage{graphicx}

\usepackage{tikz,calc,pgfplots,colortbl}
\usetikzlibrary{fit,matrix,positioning,calc,graphs,graphs.standard,
  shapes.geometric,quotes,decorations.text,decorations.markings,
  arrows,arrows.meta,backgrounds,positioning}

\ifCLASSOPTIONcompsoc
  \usepackage[nocompress]{cite}
\else
  \usepackage{cite}
\fi

\title{PHOEG Helps Obtaining Extremal Graphs}

\author{%
    \IEEEauthorblockN{Gauvain Devillez\IEEEauthorrefmark{1},
    Pierre Hauweele\IEEEauthorrefmark{1},
    Hadrien Mélot\IEEEauthorrefmark{1}
  }
  \IEEEauthorblockA{\IEEEauthorrefmark{1}
    Algorithms Lab\\
    University of Mons, Belgium\\
    Corresponding Author: pierre.hauweele@umons.ac.be
  }
}

\newtheorem*{probl}{Problem}

\newcommand*{\phoeg}{PHOEG\xspace}
\newcommand*{\pgphoeg}{PgPhoeg\xspace}
\newcommand*{\postgresql}{PostgreSQL\xspace}
\newcommand{\graphedron}{GraPHedron\xspace}
\newcommand{\transproof}{TransProof\xspace}
\newcommand{\etall}{et al.\xspace}
\newcommand{\ie}{\textit{i.e.},\xspace}
\newcommand{\eg}{\textit{e.g.},\xspace}
\newcommand{\gset}{\ensuremath{\mathcal{G}}\xspace}
\newcommand{\tset}{\ensuremath{\mathcal{T}}\xspace}
\newcommand{\extset}{\ensuremath{\mathcal{E}}\xspace}
\DeclareMathOperator{\iso}{\ensuremath{\simeq}}
\newcommand{\cpp}{\texttt{C++}\xspace}

\newcommand{\eci}{\ensuremath{\xi^c}}
\newcommand{\ecc}{\ensuremath{ecc}}
\newcommand{\degr}{\ensuremath{d}}
\newcommand{\canon}{\ensuremath{C}}
\newcommand{\dist}{\ensuremath{dist}}
\newcommand{\conv}{\ensuremath{conv}}
\newcommand{\transf}{\ensuremath{\tau}}
\newcommand{\egt}{Extremal Graph Theory\xspace}
\newcommand{\samprob}{\ensuremath{\mathcal{P}}\xspace}
\newcommand{\edge}[2]{\ensuremath{#1#2}\xspace}

\newcommand{\nodes}[1]{
  \node (title) {#1};
  \node[below of=title,node distance=1.5cm] (ct) {$\bullet$};
  \node[above left of=ct] (lt) {$\bullet$};
  \node[above right of=ct] (rt) {$\bullet$};
  \node[below of=ct] (cb) {$\bullet$};
  \node[below left of=cb] (lb) {$\bullet$};
  \node[below right of=cb] (rb) {$\bullet$};
}
\newcommand{\gedge}[3][]{\draw[#1] (#2.center) -- (#3.center);}

\tikzgraphsset{math nodes,nodes={circle, draw, inner sep=3}}

\tikzset{locatenode/.style={circle, draw, inner sep=1.5}}

\pgfplotsset{%
    colormap={gradient}{color(0cm)=(white); color(1cm)=(black)}
}

\newcommand{\drawpolytopetiny}[1]{%
    \begin{axis}[
            tiny,
        scale=2.4,
        xlabel={\footnotesize\(\xi^c\)},
        ylabel={\footnotesize\(m\)},
        ylabel near ticks,
        scatter/use mapped color={draw=black, fill=mapped color},
        colormap name=gradient,
        colorbar sampled line,
        colorbar style={
            ylabel={\footnotesize multiplicity},
        ylabel near ticks}]
\addplot[scatter, point meta=\thisrow{mult}, only marks]
        table[x=eci, y=m] {poly_#1.dat};
        \addplot[no marks] table[x=eci, y=m]
        {eci_convex_hull_#1.dat};
    \end{axis}
}

\colorlet{later}{orange!20}
\colorlet{urgent}{red}

\newtheorem{defi}{Definition}

\begin{document}
\maketitle

\begin{abstract}
    Extremal Graph Theory aims to determine bounds for graph invariants as well
as the graphs attaining those bounds.

We are currently developping PHOEG, an ecosystem of tools designed to help
researchers in Extremal Graph Theory.

It uses a big relational database of undirected graphs and works with the
convex hull of the graphs as points in the invariants space in order to
exactly obtain the extremal graphs and optimal bounds on the invariants for
some fixed parameters. The results obtained on the restricted finite class
of graphs can later be used to infer conjectures. This database also allows
us to make queries on those graphs. Once the conjecture defined, PHOEG goes
one step further by helping in the process of designing a proof guided by
successive applications of transformations from any graph to an extremal
graph. To this aim, we use a second database based on a graph data model.

The paper presents ideas and techniques used in PHOEG to assist the study
of Extremal Graph Theory.

\end{abstract}

\section{Introduction}

Graph Theory often focuses on questions about bounds for some graph
invariants. A graph invariant is a function which, given a graph \(G\)
returns a value that only depends on the structure of \(G\) --- \ie it is
invariant by isomorphism. When the bounds on these invariants are tight,
the graph realizing them are called extremal graph.

This is a specific research field in Graph Theory called \egt. A generic problem
in \egt consists in finding bounds on some invariants with respect to some
constraints. These constraints usually consist in fixing or restricting the
value of some other invariants and/or restricting the graphs to a certain class.

One of the first results in \egt is the theorem from Tur\'an~\cite{turan1941} in
1941 who determined the graphs that do not contain a clique of a given order
\(k\) and maximize the number of edges. These graphs were named the Tur\'an
graphs.

The solutions to these problems are parameterized bounds (if the value of an
invariant was fixed) and the graphs realizing those bounds. Indeed, such
extremal graphs are proofs that the bounds are tight.

These solutions obviously need to be true for all graphs respecting the given
constraints and these graphs can be numerous. An often used constraint is
to fix the order \(n\) of the graphs. But even so, there are already more
than a billion of graphs with 12 vertices.

This huge quantity of data creates a need for techniques to determine the
extremal graphs and also, to help prove their extremality.

The first project to provide these helps, called Graph,  was done by Cvetkovic
\etall in 1981~\cite{Graph}. This led later to a new version called newGRAPH by
Brankov \etall~\cite{newgraph}. But this tool was only the first of a kind and
many other tools were developped.

In 1988, Graffiti was developped by Fajtlowicz~\cite{graffiti_conjectures} and,
using heuristics and pre-computed data, was able to generate more than 7.000
conjectures in its first execution.

Later, in 2000, Caporossi and Hansen developped
AutoGraphiX~\cite{vns_autographix} which used the variable neighborhood
search metaheuristic to determine good candidates for the extremal
graphs. Digenes~\cite{digenes} (2013) uses genetic algorithms and provides
support for directed graphs.

In 2008, Mélot presented \graphedron~\cite{graphedron}. This tool differs from
the previous ones by its ideas. Indeed, rather than trying to find the extremal
graphs, \graphedron uses all the graphs up to some order in the invariant space
and then computes the convex hull of these points. The facets of the hull can
be seen as inequalities between the chosen invariants and the vertices of the
convex hull as extremal graphs. Another difference from the previous tools is
that \graphedron uses an exact approach on small graphs.

While tools such as AutoGraphiX and Graffiti have evolved over the years,
\graphedron did not. This is why we started a complete overhaul of this tool.

This successor, \phoeg, contains a set of tools aimed at speeding the testing of
ideas and helping raise new ones. It is mainly composed of a database of graphs
enabling fast queries and computations but also of a module named \transproof
whose goal is to assist finding proofs for the conjectures.

\begin{figure}[htpb]
    \centering
    \begin{tikzpicture}[
    node distance=0.4cm, scale=0.8,
    every node/.style={transform shape}]
  \draw[rounded corners] (0, 0) rectangle (10, 7);
  \draw (0, 6) -- (10, 6);
  \node at (5, 6.5) {\Large PHOEG};

  \node (Core) at (2, 5.5) {\large CoreLib};

  \node[align=center] (FGD) at (7, 5.2)
    {\large Forbidden Graph\\\large Characterization};

  \draw[rounded corners,fill=blue!20!white] (0.4, 2.6) rectangle (4.2, 0.4);
  \draw (0.4, 1.05) -- (4.2, 1.05);
  \node at (2.1, 0.7) {\footnotesize Graph Database - Neo4J};
  \node (Transproof) at (2.3, 2) {\large Transproof};
  \node[below of=Transproof,node distance=0.6cm] {\footnotesize Graph transformations};

  \draw[rounded corners,fill=red!20!white] (4.6, 4) rectangle (9.6, 0.4);
  \draw (4.6, 1.05) -- (9.6, 1.05);
  \node at (7.3, 0.7) {\footnotesize Relational Database - PostgreSQL};
  \node (IDB) at (7.1, 3.5) {\large Invariants Database};
  \node[align=center] (CHC) at (7.1, 2.2) {\large Convex Hull\\\large Calculation};
  \node[below of=CHC,node distance=0.8cm] {\footnotesize Using PostGIS};

  \draw (FGD) -- (Core);
  \draw (IDB) -- (Core);
  \draw (CHC) -- (IDB);
  \draw (Transproof) -- (Core);

  \node[below of=Core,node distance=0.6cm,fill=white] {\footnotesize Representation of graphs};
  \node[below of=Core,node distance=1cm,fill=white] {\footnotesize Invariants computation};
  \node[below of=Core,node distance=1.4cm,fill=white] {\footnotesize Various tools};
\end{tikzpicture}
    \caption{General layout of \phoeg}
    \label{fig:sumaryfig}
\end{figure}
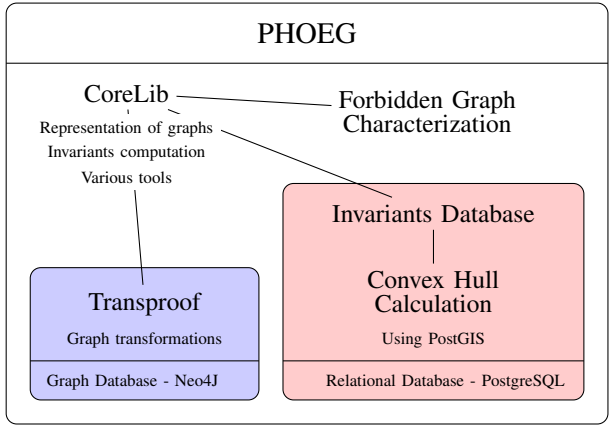

In the following sections, we present the different aspects of \phoeg and
explain some of the main ideas used to help the researcher in studying
\egt. Section~\ref{sec:corelib} describes the core library linking the
different tools together. In Section~\ref{sec:database}, we explain how the
database is built as well as how it can be used. Section~\ref{sec:convexhull}
details how the convex hull is used to generate conjectures in
\phoeg. Section~\ref{sec:fgc} describes a module helping to conjecture a
forbidden graph characterization for a specific class of graphs. Finaly,
Section~\ref{sec:transproof} presents the ideas used in \transproof to
assist the construction of a proof by transformation. An idea of the general
structure of \phoeg is presented in Figure~\ref{fig:sumaryfig}.

\section{Notations and definitions}

Common Graph Theory concepts and notations will be used. Readers that are not
familiar with these can refer to Graph Theory textbooks~\cite{diestel}. However,
we define here some specific notions and notations used in this paper.

In our work, we consider only undirected simple graphs. We note $G \iso H$ if
the two graphs $G$ and $H$ are isomorphic. In the computations, we only use one
representant for each isomorphism class called the canonical form.

\begin{defi}
    Let \gset be the set of all graphs and $G \in \gset$ be a graph. We define
    the \emph{canonical form} of $G$ (denoted by $\canon(G)$) as the result of a
    function $\canon : \gset \rightarrow \gset$ such that $\forall H \in \gset,
    \canon(H) = \canon(G) \Leftrightarrow H \iso G$.
\end{defi}

While this paper aims at presenting \phoeg and not theoretical results, we
illustrate some ideas with the following problem concerning the eccentric
connectivity index.

\begin{probl}[\samprob]
    Let \gset be the set of graphs of order $n$ and size $m$, what is the graph
    or class of graphs, among those of \gset, having the maximal eccentric
    connectivity index ?
\end{probl}

The eccentric connectivity index invariant comes from Chemical Graph Theory and
is already concerned by several theorems and conjectures~\cite{ecisurvey}. This
invariant is computed using the eccentricity and the degree of a vertex.

\begin{defi}
    Let $v$ be a vertex of a graph $G=(V,E)$ with vertex set $V$ and edge set
    $E$, the \emph{eccentricity} of $v$ ($\ecc(v)$) is the maximal distance
    between $v$ and any other vertex of $G$, \ie $\ecc(v) = \max\limits_{u \in
    V} \dist(v,u)$.
\end{defi}

\begin{defi}
    Let $G=(V,E)$ be a graph with vertex set $V$ and edge set $E$, its
    \emph{eccentric connectivity index} (denoted by $\eci(G)$) is defined as the
    sum for all vertices $v$ of the product between the eccentricity of $v$ and
    its degree, \ie $\eci(G) = \sum\limits_{v \in V} \ecc(v)\cdot \degr(v)$.
\end{defi}

In section~\ref{sec:convexhull}, The convex hull of a set of points
corresponding to graphs is used in order to produce
conjectures.

\begin{defi}
    Given a finite set of points $S = {x_0, x_1, \dots, x_n}$ in a
    $p$-dimensional space, the \emph{convex hull} of $S$ (denoted $\conv(S)$) is
    the smallest convex set containing $S$, \ie $\conv(S) =
    \left\{\sum\limits_{i=1}^{n}a_ix_i | (\forall i : a_i \geq 0) \land
    \sum\limits_{i=1}^{n}a_i = 1\right\}$. When $S$ is finite, this set forms a
    convex polytope.
\end{defi}

This polytope can be seen as an intersection of halfspaces. It can thus be
represented as a system of linear inequalities.

The \emph{facets} of the polytope are formed by intersections with halfspaces
such that none of the interior points are located on the boundaries of the
polytope. They are the "sides" of the polytope.

Section~\ref{sec:transproof} explains ideas to prove generated conjectures
with help of graph transformations.

\begin{defi}

    Let \gset be the set of all graphs, a \emph{parameterized graph
    transformation} is a function $\transf_{V,E} : \gset \rightarrow \gset$
    where $V$ is a set of vertices and $E$, a set of edges respecting some
    constraints defined by the transformation. They usually work by removing or
    adding edges and vertices given as parameters.

\end{defi}

\begin{defi}

    Given a parameterized graph transformation $\transf_{V,E}$, one can build a
    \emph{graph transformation} $\transf$ as a function that, given a graph,
    returns the set of graphs returned by $\transf_{V,E}$ for all allowed
    values of its parameters.

\end{defi}

Graph transformations can be as simple as the removal or addition of an
edge or as complex as replacing a subgraph by a new one, possibly with a
different number of vertices.

\section{Core Library}

The whole \phoeg system is built on top of a core library that models
graphs and invariants together with a set of tools. Since a desired feature
of \phoeg is to be compatible with other graph libraries --- in particular
LEMON~\cite{lemon} and the Boost Graph Library~\cite{BGL} --- it is implemented
as a \cpp header templated library.

Although \phoeg's core library supports graphs of any order, the modules
built on top of it are designed to work on small graphs (typically up to
order 10 and practically never to more than order 20). The core library
consequently provides a graph representation specialized for small graphs,
based on a binary encoding of the adjacency matrix. The resulting format is
thus quite compact and efficient.

The library also offers implementations for more than a hundreth of invariants
as well as for graph transformations. Moreover, a set of utilities functions
on graphs is provided to help defining new invariants and transformations
\eg iterators on shortest paths or cliques.

Also, thanks to the templates, one can use these functions on a custom graph
implementation as well as providing optimized algorithms for some specific
graph representation.

\label{sec:corelib}

\section{Invariants Database}

While \phoeg provides implementations for invariants and tools to write new
ones, many graph invariants come from NP-complete problems, \eg the maximal
clique or the chromatic number of graphs can take time to compute even with
optimized algorithms.

This is especially inconvenient when there are millions of graphs to consider.
This problem was tackled in \graphedron by only computing their values once
for each graph and then storing their values in files.

As invariants are constant by isomorphism, each graph is only considered once
thanks to its canonical form computed by the nauty software~\cite{nauty}.

In \phoeg, those are stored in a \postgresql database (containing at least
all the graphs up to order 10, that is more than 12 millions graphs, with
their values for each invariant). Consequently the queries are written
as SQL queries to the database. Query answering thus benefits from the
database system features such as optimizations and parallelization. An
example of a \postgresql query, using windowing functions, is given in
Figure~\ref{fig:eciextremalquery}. This query gives all the signatures (in
the graph6 format~\cite{nauty}) of all the extremal graphs for the sample
problem \samprob, together with their order, size and value of \eci.

\begin{figure}[htpb]
  \centering
  \begin{verbatim}
SELECT t.signature, t.n, t.m, t.eci
FROM (
  SELECT n.val AS n, m.val AS m,
    eci.signature, eci.val as eci,
    DENSE_RANK() OVER (
      PARTITION BY n.val, m.val
      ORDER BY eci.val DESC
    ) AS pos
  FROM num_vertices n
  JOIN num_edges m USING(signature)
  JOIN eci USING(signature)
  WHERE n.val <= 10
  ) t
WHERE t.pos = 1
ORDER BY t.n, t.m;
  \end{verbatim}
  \caption{\postgresql query obtaining all extremal graphs of order less
  or equal to 10 for \eci.}
  \label{fig:eciextremalquery}
\end{figure}

In addition, this data can be used in the computation of some invariants for
graphs of higher order. \eg the chromatic number of a graph can be computed
using the chromatic number of a subgraph. An automatic dynamic programming
module using the database as memory is currently being developed.

A work-in-progress feature to the database part is \pgphoeg, a plugin
for \postgresql adding support for graph objects. This means adding extra
datatypes for graphs plus exporting \phoeg's graph functions, invariants and
transformations to the \postgresql database. This allows for server-side
computation of the invariants and acces to graph manipulation functions
in queries, leading to less client-server context transitions and easier
writing of complex queries.

The goal of this tool is to be used by researchers in Graph Theory. As they
are not necessarily accustomed to the writing SQL queries, the addition of
a domain specific language for those kind of queries is a planned feature.
\label{sec:database}

\section{Convex Hull}\label{sec:convexhull}

In \egt, many theorems and conjectures are expressed as inequalities between
graph invariants~\cite{aouchiche,writtenonthewall}. This observation was
used by \graphedron by converting them to points whose coordinates are the
values of invariants. The convex hull of these points is then computed.

This convex hull and more precisely, its facets, provides inequalities between
the invariants used as coordinates. These correspond to bounds on their values
that can be of use to define \egt conjectures.

We note that this idea has led to several results (see e.g., \cite{MelotPhD})
including a complete answer~\cite{Christophe08} to an open problem introduced by
Ore in 1962~\cite{Ore62}. A difference from existing softwares based on
metaheuristics is that these bounds are exact for the graphs used in the convex
hull.

The Figure~\ref{fig:polytope_mult_n7} shows this polytope for the sample
problem \samprob with graphs of order 7. The points correspond to the
coordinates of the graphs in the invariants space (the size \(m\) and \eci)
and the \emph{multiplicity} counts the number of graphs satisfying the problem
constraints (connected graphs of order 7) that have the same coordinates. Those
graphs coordinates and polytope are generated from the \postgresql queries
shown in Figure~\ref{fig:polytopequery}. The convex hull is computed from
the database using PostGIS (a \postgresql plugin adding support for geometric
and geographic objects manipulation).

\begin{figure}[htbp]
  \centering
  \begin{tikzpicture}
    \drawpolytopetiny{7}
  \end{tikzpicture}
  \caption{Polytope and graph coordinates for the sample problem with \(n=7\).}
  \label{fig:polytope_mult_n7}
\end{figure}
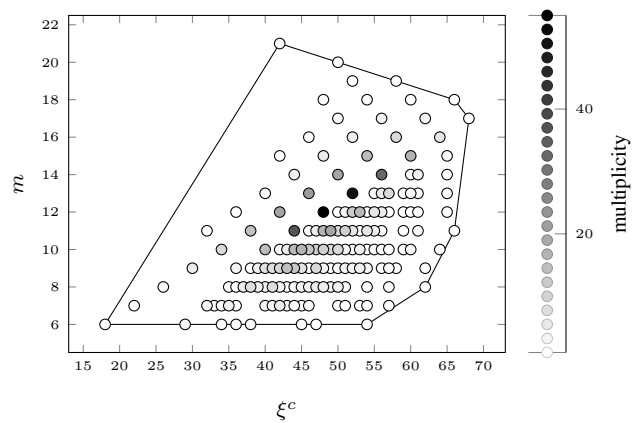

\begin{figure}[htpb]
  \centering
  \begin{verbatim}
CREATE TEMPORARY TABLE coords AS (
  SELECT num_edges.val AS m,
    eci.val AS eci, COUNT(*) AS mult
  FROM eci
    JOIN num_vertices USING(signature)
    JOIN num_edges USING(signature)
  WHERE num_vertices.val = 7
  GROUP BY m, eci
);
SELECT ST_ConvexHull(
  ST_Collect(ST_Point(m, eci)))
FROM coords;
  \end{verbatim}
  \caption{\postgresql query generating the graphs coordinates and computing
  the convex hull for the polytope.}
  \label{fig:polytopequery}
\end{figure}

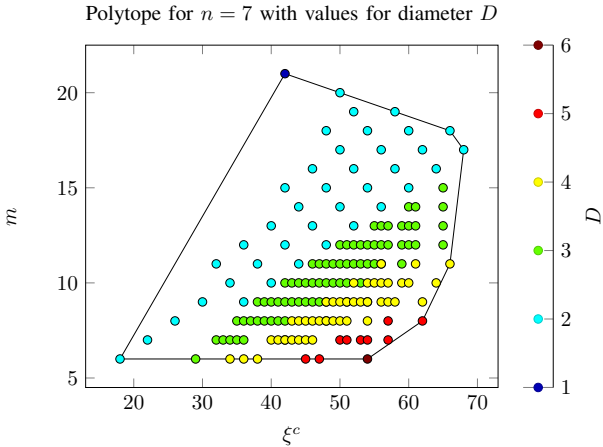
\begin{figure}[htpb]
  \centering
  \begin{tikzpicture}[scale=0.8]
    \begin{axis}[
      title={Polytope for \(n=7\) with values for diameter $D$},
      xlabel={\(\xi^c\)}, ylabel={\(m\)},
      scatter/use mapped color={draw=black, fill=mapped color},
      colormap/bluered,
      colorbar sampled line,
      colorbar style={
        ylabel={\(D\)},
        ylabel near ticks,
        samples=6}]
      \addplot[scatter, point meta=\thisrow{diam}, only marks]
        table[x=eci, y=m] {poly_diam.dat};
      \addplot[no marks] table[x=eci, y=m]
        {eci_convex_hull_7.dat};
    \end{axis}
  \end{tikzpicture}
  \caption{Coloring polytope points with values of the diameter}
  \label{fig:polytope_diam}
\end{figure}

One of the specificity of \phoeg is that it is possible to explore the inner
points of the polytope. Indeed, the access to the database and creation of
temporary tables allows to query for additional information in the polytope.
For example, Figure~\ref{fig:polytope_diam} shows a coloring of the graphs
coordinates with the maximum diameter among all the graphs with the same
coordinate.

\section{Forbidden Graph Characterization}

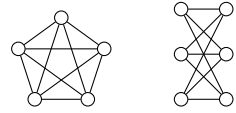
\begin{figure}[htbp]
  \centering
  \begin{tikzpicture}[scale=0.6, every node/.style={scale=0.6}]
    \graph[empty nodes, clockwise] { subgraph K_n [n=5] };
  \end{tikzpicture}
  \qquad
  \begin{tikzpicture}[scale=0.6, every node/.style={scale=0.6}]
    \graph[empty nodes] { subgraph K_nm [n=3, m=3] };
  \end{tikzpicture}
  \caption{Obstruction set of the Kuratowski's theorem.}
  \label{fig:kuratowski_obstruct}
\end{figure}

In \egt, and in Graph Theory in general, classes of graphs are often described
by means of a forbidden graph characterization. Such a characterization of a
class \(\mathcal{G}\) of graphs is given by an obstruction set \(\mathcal{O}\)
containing the forbidden graphs. A graph \(G\) is a member of \(\mathcal{G}\)
if and only if it has no element of \(\mathcal{O}\) as substructure (\eg
induced subgraph, graph minor). A classical example of such a characterization
is given by the Kuratowski's theorem~\cite{kuratowski1930}. It states that a
finite graph is planar if and only if it has no \(K_5\) (the complete graph of
order 5) nor \(K_{3,3}\) (the complete bipartite graph of order 6, see
Figure~\ref{fig:kuratowski_obstruct}) as
topological minor.

\phoeg provides a tool adressing this matter. The substructure relations define
a preorder. Given a finite class of graphs or a class membership function
and a specific substructure relation, \phoeg computes the minimal graphs
(not) in the class for this relation, using, \eg the VF2 algorithm~\cite{vf2}
for the subgraph (iso/mono)morphism relation. The output set of
minimal graphs provides a proposed obstruction set for the forbidden graph
characterization of the input class.
\label{sec:fgc}

\section{Transproof}\label{sec:transproof}

After obtaining conjectures, one needs to prove them. Let \gset be the set of
graphs concerned by the conjecture and $\extset \subseteq \gset$, the set of
extremal graphs, a common technique is a proof by transformation. This kind of
proof works by defining a set of graph transformations (denoted by \tset) such
that, for any graph in $\gset \backslash \extset$, there is a transformation
returning a new graph, non isomorphic to the previous one and whose value for
the studied invariant is closer to the conjectured bound. Incidentally, there
exists a proof by transformation of the Tur\'an
theorem~\cite[p.272-273]{proofsbook}.

One of the most difficult parts of such proofs is finding good transformations.
Actually, one not only wants correct transformations but also simple
transformations to simplify the proof and as few as possible to avoid long and
repetitive proofs (an example with more than 40 transformations can be seen
in~\cite{alain42}). We define the simplicity of a transformation as the number
of elements of the graph (vertices and edges) touched by the transformation.

Another way to represent these proofs is on the form of a directed graph. The
vertices of this graph are the graphs concerned by the conjecture (\gset)
and an arc from vertex $A$ to vertex $B$ means that there is a transformation
from the graph $A$ to the graph $B$. We call this graph, the \emph{metagraph
of transformations}. With this point of view, a proof by transformation
is correct if the metagraph is acyclic and all its sinks (vertices with no
exiting arcs) are extremal graphs.

\begin{figure}[!ht]
    \centering
    \includegraphics[width=2.5in]{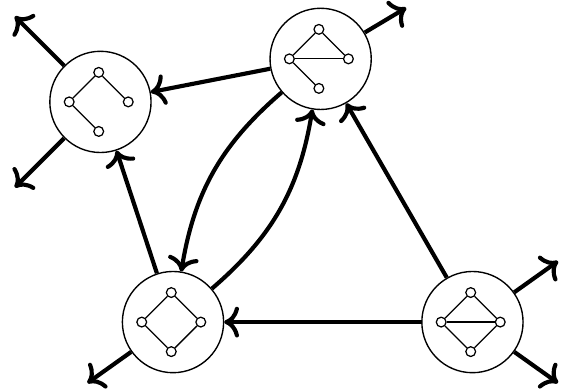}
    \caption{Metagraph example for edge deletion and rotation.}
    \label{fig:metagraph}
\end{figure}

\begin{defi}

    Let \gset be a set of graphs and \tset be a set of graph transformations.
    Let $M$ be the directed graph with vertex set \gset and arc set ${ (G,U) \in
    \gset\times\gset | \exists \tau \in \tset \land \exists H \in \tau(G), H
    \iso U}$. We call this graph the \emph{metagraph of transformations} for the
    given graph set \gset and transformation set \tset. An example is given in
    Figure~\ref{fig:metagraph}.

\end{defi}

This idea is exploited in the \transproof module. The metagraph is pre-computed,
for a given set of graphs and transformations, and stored inside a database
using the graph data model (currently Neo4j~\cite{neo4j}). This specific NoSQL
database makes queries on the structure of the metagraph quite efficient since
there is no need to perform joins between tables to find a path.

This provides a basis on which tools can be built to allow study of the
metagraph. It can be used to test the efficiency of a transformation if used in
a proof but also, to help refine them if they are not correct. That mechanism
can also be exploited in building heuristic tools to provide good candidates
for such proof by making the evaluation of a transformation faster than
recomputing it everytime.

However, as the number of potential arcs can be quadratic in the number of
graphs, the database grows in size exponentially making it impossible to compute
results for all graphs and even less for all possible transformations. This thus
brings need to resort to heuristics in order to construct proofs by
transformation.

The idea to try and overcome these limitations is that a transformation can be
described as a sequence of simple transformations (transformations with little
impact on the graph they are applied to). For example, a commonly used
transformation is the rotation.

\begin{defi}

    Let $G = (V,E)$ be a graph. Let $a,b,c$ be vertices in $V$ such that
    $\edge{a}{b} \in E$ and $\edge{a}{c} \not\in E$. The \emph{rotation} on $G$
    using the vertices $a,b,c$ is a parameterized graph transformation that
    removes edge \edge{a}{b} and adds edge \edge{a}{c}.

\end{defi}

A rotation thus consists in removing an edge and adding another one. Both these
transformations are really simple ones. We can thus only define some simple
transformations to be precomputed and stored in the database and use them to
generate other more complex transformations by chaining them and adding
constraints.

\begin{figure}[!ht]
    \centering
    \includegraphics[width=2.5in]{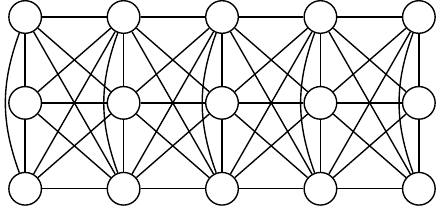}
    \caption{Example where moving a vertex is necessary to increase the diameter}
    \label{fig:incdiam}
\end{figure}

The choice of this basis of transformations requires to be able to generate
all transformations from a subset of the simple ones. As an example, one can
consider a path of length $l$ where each vertex is replaced by a clique of
a given size $k$ and where the clique replacing the vertex in position $i$
is fully joined with the cliques replacing the vertices in position $i-1$
and $i+1$ if they exist.  Figure~\ref{fig:incdiam} represents such a graph
for $l = 5$ and $k=3$. This graph has diameter $5$ and one has to remove at
least $k$ edges to increase it.

This means that we need not only transformations about edges but also about
vertices. \eg moving a vertex (removing its adjacent edges and adding new ones)
means removing and adding a non fixed number of edges depending of the structure
of the graph. We also need a way to specify how these edges will be added. We
should thus add different transformations based on the ways one can add a vertex
to the graph.

With this basis of transformations, we need only to compute simple
transformations and store them inside the graph database. This data can then
be exploited by queries to the database for more complex transformations.

To this extent, a specific language is being developped where a statement
consists of a list of transformations to apply to some parts of the graph
and, for each of these transformations, a set of constraints potentially
empty. We can thus consider more complex transformations but, for the same
reasons as evoked in Section \ref{sec:database},
we are still limited to small graphs.

\begin{figure}[htpb]
    \centering
    \def\edgeaddthickness{ultra thick}
\begin{tikzpicture}[
    scale=0.7,
    every node/.style={transform shape}]
  \begin{scope}
    \nodes{}
    \gedge{lt}{ct}
    \gedge{rt}{ct}
    \gedge{ct}{cb}
    \gedge{lb}{cb}
    \gedge{rb}{lb}
  \end{scope}

  \begin{scope}[xshift=3cm]
    \nodes{Removing}
    \gedge[dashed]{lt}{ct}
    \gedge{rt}{ct}
    \gedge{ct}{cb}
    \gedge{lb}{cb}
    \gedge{rb}{lb}
  \end{scope}

  \begin{scope}[xshift=6cm]
    \nodes{Adding}
    \gedge{lt}{ct}
    \gedge{rt}{ct}
    \gedge{ct}{cb}
    \gedge{lb}{cb}
    \gedge{rb}{lb}
    \gedge[\edgeaddthickness]{lt}{lb}
  \end{scope}

  \begin{scope}[yshift=-4.5cm]
    \nodes{Rotation}
    \gedge[dashed]{lt}{ct}
    \gedge{rt}{ct}
    \gedge{ct}{cb}
    \gedge{lb}{cb}
    \gedge{rb}{lb}
    \gedge[\edgeaddthickness]{ct}{lb}
    \node[fill=white] at (lt) {$b$};
    \node[fill=white] at (ct) {$a$};
    \node[fill=white] at (lb) {$c$};
  \end{scope}

  \begin{scope}[yshift=-4.5cm,xshift=3cm]
    \nodes{Moving}
    \gedge{lt}{ct}
    \gedge[dashed]{rt}{ct}
    \gedge{ct}{cb}
    \gedge{lb}{cb}
    \gedge{rb}{lb}
    \gedge[\edgeaddthickness]{lt}{lb}
  \end{scope}

  \begin{scope}[yshift=-4.5cm,xshift=6cm]
    \nodes{Detour}
    \gedge[dashed]{lt}{ct}
    \gedge{rt}{ct}
    \gedge{ct}{cb}
    \gedge{lb}{cb}
    \gedge{rb}{lb}
    \gedge[\edgeaddthickness]{lt}{lb}
    \gedge[\edgeaddthickness]{lb}{ct}
    \node[fill=white] at (lt) {$a$};
    \node[fill=white] at (ct) {$b$};
    \node[fill=white] at (lb) {$c$};
  \end{scope}

  \begin{scope}[yshift=-9cm,xshift=1.5cm]
    \nodes{Shortcut}
    \gedge{lt}{ct}
    \gedge{rt}{ct}
    \gedge[dashed]{ct}{cb}
    \gedge[dashed]{lb}{cb}
    \gedge{rb}{lb}
    \gedge[\edgeaddthickness]{lb}{ct}
    \node[fill=white] at (ct) {$a$};
    \node[fill=white] at (lb) {$b$};
    \node[fill=white] at (cb) {$c$};
  \end{scope}

  \begin{scope}[yshift=-9cm,xshift=4.5cm]
    \nodes{Two-opt}
    \gedge{lt}{ct}
    \gedge[dashed]{rt}{ct}
    \gedge{ct}{cb}
    \gedge{lb}{cb}
    \gedge[dashed]{rb}{lb}
    \gedge[\edgeaddthickness]{ct}{lb}
    \gedge[\edgeaddthickness]{rt}{rb}
    \node[fill=white] at (ct) {$a$};
    \node[fill=white] at (rt) {$b$};
    \node[fill=white] at (lb) {$c$};
    \node[fill=white] at (rb) {$d$};
  \end{scope}
\end{tikzpicture}
    \caption{Illustration of the transformations stored in the graph database}
    \label{fig:transfos}
\end{figure}
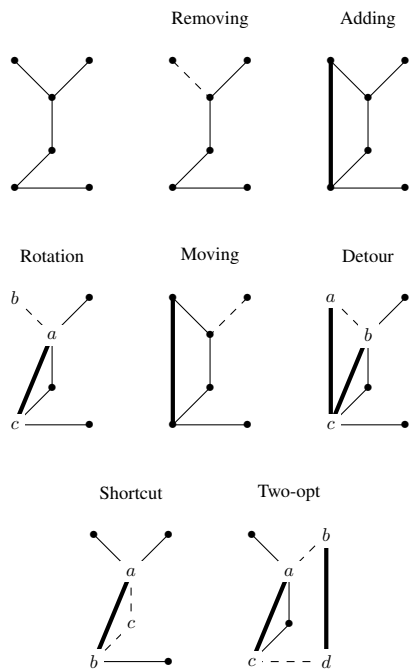

At the time of writing, the graph database contains eight simple
transformations, illustrated in Figure~\ref{fig:transfos} :

\begin{IEEEitemize}

    \item removing an edge.

    \item adding an edge.

    \item rotation

    \item moving an edge.

    \item detour : given \edge{a}{b} in $E$ and $c$ in $V$ not adjactent to $a$
        or $b$, we replace \edge{a}{b} by a path of length 2 joining $a$ to
        $b$ by going through $c$.

    \item shortcut : given a path of length 2 joining $a$ to $b$ by going
        through $c$, we replace that path by a single edge \edge{a}{b}.

    \item two-opt : given \edge{a}{b} and \edge{c}{d} in $E$ such that
        \edge{a}{c} and $\edge{b}{d} \not \in E$, we replace \edge{a}{b} and
        \edge{c}{d} by the \edge{a}{c} and \edge{b}{d}.

    \item slide : given \edge{a}{b} in E $c$ in $V$ such that $\edge{a}{c} \not
        \in E$ and $\edge{b}{c} \in E$, we replace \edge{a}{b} by \edge{a}{c}.
        This is actually a rotation on $a,b$ and $c$ that conserve connectivity.

\end{IEEEitemize}

They are precomputed on all the graphs up to order 9. The number of arcs in the
database for the different orders can be seen in Table~\ref{tab:numtransfos}.
This data is currently being used to assist finding proofs for several
conjectures.

\begin{table}[!ht]
    \centering
    \caption{Number of arcs in the metagraph for graph orders from 2 to 9.}
    \label{tab:numtransfos}
    \begin{tabular}{crr}
        \hline
        order & \# graphs &\# arcs \\
        \hline
        1 & 1 & 0 \\
        2 & 2 & 4 \\
        3 & 4 & 36 \\
        4 & 11 & 362 \\
        5 & 34 & 3~188 \\
        6 & 156 & 34~376 \\
        7 & 1~044 & 468~936 \\
        8 & 12~346 & 10~143~824 \\
        9 & 274~668 & 380~814~904 \\
        \hline
    \end{tabular}
\end{table}







\section{Conclusion and future work}

We explained how \phoeg exploits its databases to help finding conjectures in
\egt but also characterize classes of graphs \emph{via} forbidden subgraphs and
construct proofs by transformations.

\phoeg is already used for some open problems in \egt but there is always room
for improvement.

Indeed, the PostgreSQL database can be used in a form of dynamic programming to
compute invariants but also to generate graphs. This asks for ways to keep track
of the partially generated classes to avoid recomputing the same graph twice.

The graph database can also be improved by introducing filters removing
symmetries among transformations. This could be done by the exploitation of
graph automorphism and could greatly reduce the number of arcs of the metagraph.



\bibliographystyle{IEEEtran}
\bibliography{biblio}

\end{document}